\documentclass[prl,amsmath,showpacs,twocolumn]{revtex4}
\usepackage{graphicx}
\usepackage{multirow}
\usepackage{amssymb}

\begin{document}

\title{{\sl Ab initio} Evidence for Giant Magnetoelectric Responses
  Driven by Structural Softness}

\author{Jacek C. Wojde\l\ and Jorge \'I\~niguez}

\affiliation{Institut de Ci\`encia de Materials de Barcelona
(ICMAB-CSIC), Campus UAB, 08193 Bellaterra, Spain}

\begin{abstract}
  We show that inducing {\sl structural softness} in regular
  magnetoelectric (ME) multiferroics -- i.e., tuning the materials to
  make their structure strongly reactive to applied fields -- makes it
  possible to obtain very large ME effects. We present illustrative
  first-principles results for BiFeO$_3$ thin films.
\end{abstract}

\pacs{75.80.+q, 77.55.Nv, 77.80.B-, 71.15.Mb}






\maketitle

Magnetoelectric (ME) multiferroics present coupled electric and
magnetic orders~\cite{fiebig05}, which could lead to novel
devices. However, finding multiferroic compounds that display large ME
effects at room temperature ($T_{\rm r}$) is proving a major
challenge, and so far we know only one candidate: the perovskite
BiFeO$_3$ (BFO)~\cite{catalan09}. The difficulties begin with the
scarcity of ferroelectrics (FEs) magnetically ordered at $T_{\rm
  r}$~\cite{filippetti02}. Moreover, the desired materials must
present large ME couplings and be good insulators, as needed for many
envisioned applications. Here we propose a general and robust design
strategy to satisfy all such requirements. Our approach -- i.e.,
inducing {\sl structural softness} to obtain strong ME effects -- is
illustrated with first-principles results for BFO films.

{\sl Materials-design strategies}.-- Let us recall the formalism
recently introduced for computing the ME response~\cite{wojdel09}. The
linear ME tensor can be written as
\begin{equation}
\label{eq:totalalpha}
\boldsymbol{\alpha} = \boldsymbol{\alpha}^{\rm el} + \Omega_0^{-1}
\boldsymbol{Z}^{T} \boldsymbol{K}^{-1} \boldsymbol{\zeta}
+ \boldsymbol{e} \boldsymbol{C}^{-1} \boldsymbol{h}
\, ,
\end{equation}
where $\Omega_0$ is the cell volume and contributions from both spin
and orbital magnetism are in principle
considered. $\boldsymbol{\alpha}$ consists of purely electronic
($\boldsymbol{\alpha}^{\rm el}$) and lattice-mediated parts, the
latter being split in frozen-cell ($\Omega_0^{-1} \boldsymbol{Z}^{T}
\boldsymbol{K}^{-1} \boldsymbol{\zeta}$) and strain-mediated
($\boldsymbol{e} \boldsymbol{C}^{-1} \boldsymbol{h}$) terms. Let us
discuss how to engineer them.

The mechanisms contributing to $\boldsymbol{\alpha}^{\rm el}$ share
the basic feature that the energy cost for polarizing the electrons is
roughly given by the band gap. Hence, having a relatively small gap
seems necessary to obtain very large effects, but this is detrimental
to the insulating character of the material. Electronic FEs like
LuFe$_2$O$_4$~\cite{lufe2o4} seem to fit this description, and will
probably display the largest effects achievable based on electronic
mechanisms.

The lattice-mediated part of the response is proportional to the
fundamental electro- and magneto-structural couplings, i.e., to the
changes of polarization and magnetization caused by atomic
displacements (quantified, respectively, by the Born effective charges
$\boldsymbol{Z}$ and the {\sl magnetic strengths}
$\boldsymbol{\zeta}$) or strain (quantified, respectively, by the
piezoelectric and piezomagnetic stress tensors $\boldsymbol{e}$ and
$\boldsymbol{h}$). The possibility to enhance the electro-structural
couplings $\boldsymbol{Z}$ and $\boldsymbol{e}$ seems unpromising, as
their magnitude is basically controlled by the nominal ionization
charges and they are already anomalously large in most FE
perovskites~\cite{filippetti02,zhong94,neaton05}. To increase the
magneto-structural couplings $\boldsymbol{\zeta}$ or $\boldsymbol{h}$,
one would typically need to use heavy magnetic species presenting
strong spin-orbit effects. However, transition metals with relatively
extended 4$d$ and 5$d$ orbitals tend to result in metallic
states~\cite{lezaic}. The insulating character can be obtained by
using lanthanides, but at the expense of weak magnetic interactions
associated to the localized 4$f$ orbitals. Note that
exchange-striction mechanisms have also been studied~\cite{delaney09},
but the ME couplings obtained so far are relatively small. Finally, a
large orbital contribution to the ME response will typically require
strong spin-orbit effects and {\sl unquenched} magnetic moments, which
seems generally incompatible with a robust insulating
character~\cite{fn:orbital}; we thus seem restricted to spin
magnetism.

We are thus left with the tensors that quantify the energy cost
associated to structural distortions, be it atomic displacements (the
force-constant matrix $\boldsymbol{K}$) or cell strains (the elastic
constant tensor $\boldsymbol{C}$). Interestingly, many structural
phase transitions are associated to the {\sl softening} of the
lattice, i.e., to a vanishingly small eigenvalue of $\boldsymbol{K}$
or $\boldsymbol{C}$. This is the case of soft-mode FEs, for which many
strategies to tune the materials properties -- by means of epitaxial
strain~\cite{zeches09}, chemical substitution~\cite{noheda99,kan10},
etc. -- have been demonstrated. Such property tuning usually implies
controlling $\boldsymbol{K}$ or $\boldsymbol{C}$: For example, the
large piezoelectric-strain response
($\boldsymbol{d}=\boldsymbol{e}\boldsymbol{C}^{-1}$) of
PbZr$_{1-x}$Ti$_{x}$O$_3$ (PZT) results from the softening of
$\boldsymbol{C}$; in SrTiO$_3$, the softness of $\boldsymbol{K}$
results in a large dielectric response
($\boldsymbol{\chi}\sim\boldsymbol{Z}^{T}\boldsymbol{K}^{-1}\boldsymbol{Z}$). According
to Eq.~(\ref{eq:totalalpha}), structural softness will also result in
large values of $\boldsymbol{\alpha}$, the enhancement being
essentially independent of the details of the electro- and
magneto-structural couplings. Hence, we can directly borrow ideas from
the field of FEs and try to obtain large ME responses driven by
structural softness.

{\sl Test case: BiFeO$_3$ films}.-- BFO is ideal to test this design
strategy. Recent studies \cite{zeches09,bea09} show that under
compressive epitaxial strain (001)-BFO films undergo a structural
transition involving two FE phases. This suggests there is strain
range within which BFO films might be structurally soft, thus having
the potential to display large responses. In addition, a recent
theoretical work \cite{wojdel09} shows that the ME response of BFO is
dominated by low-energy distortions involving the Bi and O
atoms. Further, it was found that the electro- and magneto-structural
couplings are not particularly large in BFO. Thus, BFO poses a
demanding test for our structural-softness strategy; if it works for
this compound, we will have evidence it may constitute a robust,
universal route to enhance the ME response of multiferroics.

For our simulations we used the so-called ``LDA+$U$'' approach to
Density Functional Theory, the details being as in
Ref.~\onlinecite{wojdel09}. We used a 10-atom cell given in a
Cartesian setting by:
$\boldsymbol{a}_{1}$=$(\delta_{2},a+\delta_{1},c)$,
$\boldsymbol{a}_{2}$=$(a+\delta_{1},\delta_{2},c)$, and
$\boldsymbol{a}_{3}$=$(a,a,0)$. This cell is compatible with the
atomic structure of the two FE phases of interest~\cite{zeches09}: an
``{\sl R}~phase'' that is similar to the rhombohedral $R3c$ phase of
bulk BFO, and a ``{\sl T}~phase'' that resembles the tetragonal $P4mm$
phase of BiCoO$_3$. More precisely, this cell allows for general FE
and anti-ferrodistortive (AFD) distortions associated, respectively,
to the $\Gamma_{4}^{-}$ and $R_{4}^{+}$ representations of the
reference space group $Pm\bar{3}m$. Because of the epitaxial mismatch
in the (001)$_{c}$ Cartesian plane, these distortions will typically
split into out-of-plane ($\Gamma_{4,z}^{-}$ and $R_{4,z}^{+}$) and
in-plane ($\Gamma_{4,x}^{-}$=$\Gamma_{4,y}^{-}$ and
$R_{4,x}^{+}$=$R_{4,y}^{+}$) components, reflecting a monoclinic $Cc$
symmetry. Note that the AFD distortions correspond to the O$_6$
octahedra {\sl rotations} around [001]$_{c}$ ($R_{4,z}^{+}$) and {\sl
  tilts} around [110]$_{c}$ ($R_{4,x}^{+}$=$R_{4,y}^{+}$) usually
discussed in the literature. Finally, the chosen cell is compatible
with the magnetic order of both FE phases, which we checked to be
G-type anti-ferromagnetic (G-AFM) in the strain range of
interest~\cite{fn:magorder}. For each value of the epitaxial strain
$\epsilon$, we ran constrained relaxations to find the equilibrium
structure, determined the magnetic easy axis, and computed the
response properties as in
Ref.~\onlinecite{wojdel09}~\cite{fn:extradets}.

\begin{figure}[t!]
\begin{centering}
\includegraphics[width=0.90\columnwidth]{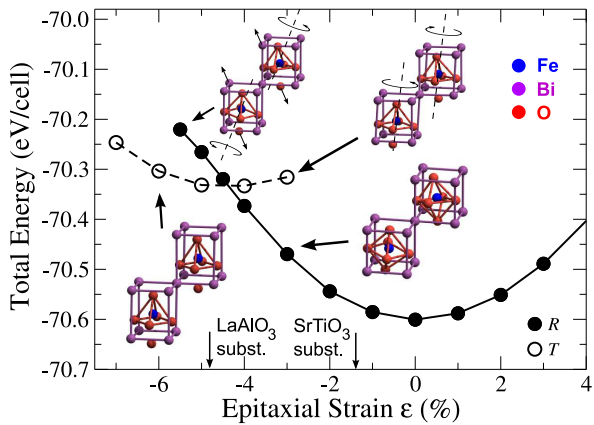}
\end{centering}
\vskip -2mm
\caption{Energy vs. epitaxial strain for the {\sl R} and {\sl T}
  phases. Computed soft modes are sketched (see text).}
\label{fig1}
\end{figure}

Figure~1 shows the computed $E(\epsilon)$ curves for the {\sl R} and
{\sl T} phases. At variance with the study of
Ref.~\onlinecite{zeches09}, we were able to track the two phases up to
their (meta)stability limits, thus locating the boundaries of the
region within which they can co-exist. We found that the {\sl R} phase
can occur up to an epitaxial compression of about $-$6\%, and the {\sl
  T} phase is predicted to occur for compressions above $-$3\%. Beyond
its (meta)stability limit, the {\sl R} (resp.~{\sl T}) phase relaxes
into the {\sl T} (resp.~{\sl R}) phase in our simulations. We thus
confirm the prediction of a first-order isosymmetric transition
between the {\sl R} and {\sl T} phases~\cite{zeches09}, with an ideal
transition point at $\epsilon$$\approx$$-$4.4\% and an hysteretic
behavior confined within the $-$6\%~to~$-$3\% region.

Our simulations allowed us to monitor the softening of the structural modes
that result in the destabilization of the FE phases. Interestingly, both soft
modes (sketched in Fig.~1) have a strong AFD character~\cite{fn:modesym}: For
the {\sl T}~phase, 93\% of the mode eigenvector is AFD in nature (83\%
rotation and 10\% tilt), and this mode accounts for 75\% of the {\sl
  T}-to-{\sl R} transformation. For the {\sl R}~phase, 67\% of the mode
eigenvector is AFD (41\% rotation and 26\% tilt) and the FE component reaches
18\% (12\% and 6\% of out-of-plane and in-plane distortions, respectively);
this mode accounts for 97\% of the {\sl R}-to-{\sl T} transition. Our results
show that, while it is correct to describe the {\sl R}$\leftrightarrow${\sl T}
transitions as FE-to-FE, the primary order parameter for them is actually AFD
in nature. Note that this is consistent with the fact that the AFD distortion
is the strongest structural instability of bulk BFO.

\begin{figure}[t!]
\begin{centering}
\includegraphics[width=0.85\columnwidth]{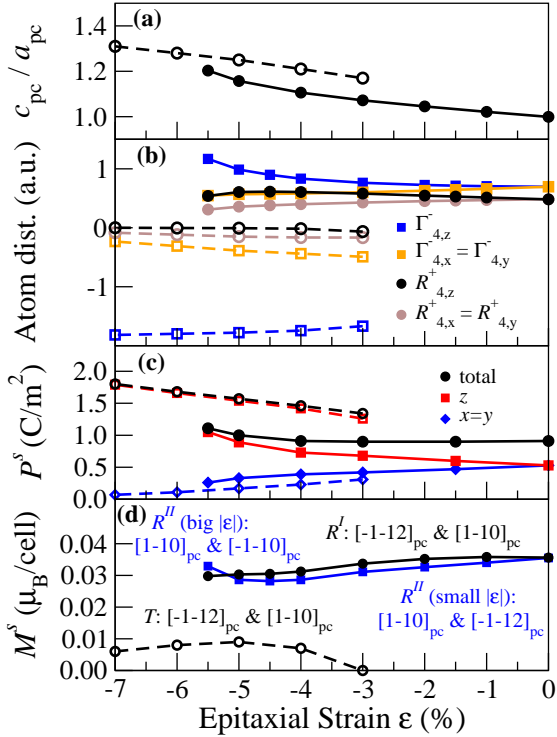}
\end{centering}
\vskip -2mm
\caption{Properties of the {\sl R} (filled symbols, solid lines) and
  {\sl T} (open, dashed) phases. Panel~(a): Aspect ratio of the
  pseudo-cubic (pc) cell associated to relaxed structures.  Panel~(b):
  Atomic structure given as a set of distortions (in arbitrary units)
  of prototype $Pm\bar{3}m$ phase as obtained with the ISODISPLACE
  software~\protect\cite{isodisplace}; data condensed by adding up
  contributions from isosymmetric modes; small contributions not
  associated to $\Gamma_{4}^{-}$ or $R_{4}^{+}$ not shown; for
  clarity, {\sl T}-phase results were chosen to be
  negative. Panels~(c) and (d): Spontaneous polarization and
  magnetization. Easy axis \& $\boldsymbol{\cal M}^{s}$ directions
  indicated in (d); note that two {\sl R} phases were considered (see
  text).\label{fig2}}
\end{figure}

Figure~2 summarizes how the properties of the {\sl R} and {\sl T}
phases change as a function of $\epsilon$. The results for the
structure (panels a and b) and polarization (panel~c) show a
tetragonalization of both phases as the epitaxial compression
grows. The main differences between them are apparent: The {\sl
  R}~phase presents much stronger O$_6$ rotations and the {\sl
  T}~phase is characterized by a very large out-of-plane polar
distortion. We can compare our results at $\epsilon$=$-$4.8\%
(LaAlO$_3$ substrate) with the experimental characterization of the
{\sl T}~phase. The values measured at $T_{\rm room}$ are~\cite{bea09}:
$c/a$=1.23, ${\cal P}_{z}^{s}$=0.75~C/m$^2$, and $d_{\rm
  eff}$=30~pC/N. We obtained $c/a$=1.23, ${\cal
  P}_{z}^{s}$=1.5~C/m$^2$, and $d_{\rm 33}$=18~pC/N for the {\sl
  T}~phase, and $c/a$=1.14, ${\cal P}_{z}^{s}$=0.8~C/m$^2$, and
$d_{\rm 33}$=87~pC/N for the {\sl R}~phase. As noted
before~\cite{bea09}, the measured ${\cal P}_{z}^{s}$ is surprisingly
small; additionally, we found it is in (probably fortuitous) agreement
with our result for the {\sl R}~phase.

Figure~2d summarizes our results for the magnetic
structure~\cite{fn:magsym}. We found that the magnetic ground state of the
{\sl R}~phase changes from {\sl R}$^{I}$ to {\sl R}$^{II}$ (notation from
Fig.~2d) for compressions above $-$5\%. The computed spontaneous magnetization
$\boldsymbol{\cal M}^{s}$ is significant only for the {\sl R}~phases, and
tends to decrease for increasing compression.  This probably reflects the fact
that the AFD distortion -- which tends to diminish with increasing compression
for the {\sl R} phases and is small for the {\sl T} phase -- is necessary for
the spin canting to exist in the kind of G-AFM ground states we
obtained~\cite{ederer05}.

\begin{figure}[t!]
\begin{centering}
  \includegraphics[width=0.85\columnwidth]{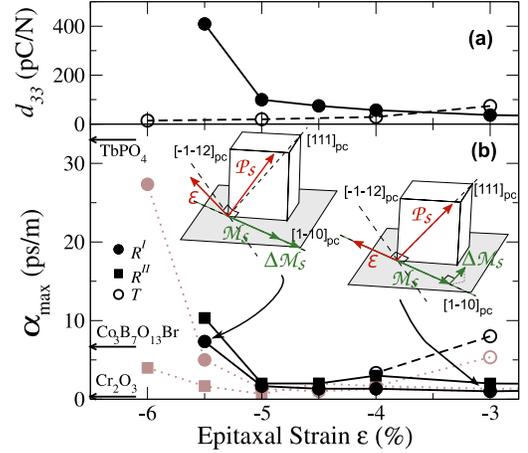}
\end{centering}
\vskip -2mm
\caption{Piezoelectric (a) and ME (b) responses of the {\sl R} (filled
  symbols, solid lines) and {\sl T} (open, dashed) phases. Panel~(b):
  $\alpha_{\rm max}$ is the square root of the largest eigenvalue of the
  quartic form $\boldsymbol{\alpha}^{T}\boldsymbol{\alpha}$. Frozen-cell
  (light colored) and full (dark) lattice responses are given. Insets sketch
  $\alpha_{\rm max}$ response of {\sl R}$^{I}$ (see text) at small and strong
  epitaxial compression. $\alpha_{\rm max}$ for three representative MEs are
  indicated (data taken from Ref.~\protect\onlinecite{borovik06}; note 1~ps/m
  = 3$\times$10$^{-4}$ Gaussian units).}
\end{figure}

Finally, Fig.~3 shows two representative responses: piezoelectric
($d_{33}$) and magnetoelectric ($\alpha_{\rm max}$, as defined in the
figure caption). For all phases, the responses increase as we approach
the stability limit. We explicitly checked this enhancement is driven
by structural softness: a mode-by-mode decomposition \cite{wojdel09}
of the frozen-cell ME response showed that the instability modes
become dominant near the correspnding co-existence boundary. This is
most apparent in the frozen-cell ME response of the {\sl R}~phases at
$\epsilon$=$-6$\%~\cite{fn:minus6}, which displays a large enhancement
associated to a very soft $\boldsymbol{K}$-eigenvalue of
0.3~eV/\AA$^2$ (to be compared with a lowest-lying
$\boldsymbol{K}$-eigenvalue of 3.2~eV/\AA$^2$ for bulk BFO). Note that
a significant part of the total response is strain-mediated,
especially in the case of {\sl R}$^{II}$.

The dominant AFD character of the soft modes is key to understand the computed
responses. Firstly, the weakly polar soft modes couple weakly with an applied
electric field. Hence, the instability modes need to become very soft to
dominate the responses, and the softness-driven enhancement (strictly
speaking, {\sl divergence}) is confined to relatively narrow regions of
epitaxial strain close to the co-existence boundaries. To better appreciate
this weakly polar nature, note that the electric polarity~\cite{fn:polarities}
of the {\sl R}~phase soft mode has a largest component of about 2.2$|e|$, $e$
being the electron charge, while strongly polar modes for the same state reach
values of 7.4$|e|$. For the {\sl T}~phase, the largest component of the soft
mode dielectric polarity varely reaches 1$|e|$.

Secondly, the marked AFD character of the soft mode of the {\sl R}
phase determines the nature of the strongest ME effects obtained,
which correspond to {\sl R}$^{I}$. The results are sketched in
Fig.~3b: For small compressions, $\alpha_{\rm max}$ is associated to
the development of a magnetization $\Delta \boldsymbol{\cal M}$
perpendicular to $\boldsymbol{\cal M}^{s}$; the dominant modes do not
present any AFD component. For strong compressions, $\alpha_{\rm max}$
corresponds to a change in the magnitude (not direction) of ${\cal
  M}^{s}$. That response is dominated by a weakly-polar soft mode with
a large AFD component; its effect in the magnitude of the canted
magnetic moment is thus natural, as such a canting was shown to be
proportional of the amplitude of the O$_6$
rotations~\cite{ederer05}. Finally, note that the computed soft modes
do not display particularly large magnetic polarities
$\boldsymbol{p}^{m}$~\cite{fn:polarities}: a largest value of
2$\times$10$^{-3}$~Bohr magnetons ($\mu_{\rm B}$) was obtained for
{\sl R}$^{I}$, while values of about 6$\times$10$^{-3}$$\mu_{\rm B}$
were obtained for the more magnetically active modes. We observed the
largest $\boldsymbol{p}^{m}$'s tend to correspond to modes
characterized by Fe displacements.

We have thus shown that structural softness results in a large
enhancement of the ME response of the BFO films, even if the somewhat
inappropriate character of the observed soft modes (i.e., their
relatively small electric and magnetic polarities) is detrimental to
the effect.  To put our results in perspective, we have indicated in
Fig.~3b the measured ME response of several representative materials
(see Table~1.5.8.2 of Ref.~\onlinecite{borovik06}): TbPO$_{4}$
(strongest single-phase magnetoelectric), Co$_{3}$B$_{7}$O$_{13}$Br
(strongest transition-metal magnetoelectric), and Cr$_2$O$_3$ (a
material with typically small $\boldsymbol{\alpha}$). Note that the
indicated $\alpha$'s correspond to temperatures slightly below the
magnetic ordering transition, where the ME effect is strongly
enhanced. Remarkably, the responses of the {\sl R} and {\sl T} phases
of BFO films, which we computed at $T$=0~K, are comparable to these
largest of $\alpha$'s for a significant range of epitaxial strains.

As to the practical implications of our results, one should first note
that, for any value of $\epsilon$, the predicted stable phase is not
soft enough to have a strong ME response. Nevertheless, as regards the
real BFO films at $T_{\rm room}$, the large reactivity to electric
fields observed by Zeches {\sl et al}.~\cite{zeches09} in the samples
presenting a {\sl mixed} {\sl R}--{\sl T} state is clearly suggestive
of the structural softness discussed here, and might thus be
accompanied by large ME effects. Additionally, let us note that
BFO-based solid solutions may provide a more convenient alternative to
soften the BFO lattice. Indeed, enhanced electromechanical responses
have been observed in compounds in which a rare earth substitutes for
Bi~\cite{kan10}, and the substitution of Fe by Co~\cite{azuma08} has
been shown to result in a monoclinic phase that acts as a {\sl
  structrual bridge} between BiCoO$_3$'s {\sl T} and BiFeO$_3$'s {\sl
  R} phases, as in strong piezoelectric PZT~\cite{noheda99}. Our
results clearly suggest that such compounds will present very large ME
respones.

In summary, we have shown that structural softness constitutes a
robust mechanism leading to large ME responses. Our simulations for
BFO films show that, even if this material is in some ways
inappropriate for the engineering of the ME response, the induction of
structural softness results in effects comparable with the greatest
ones ever measured for single-phase compounds. We hope our results
will motivate the experimental exploration of this strategy either in
BFO-based systems or in other, better suited, compounds where the soft
structural modes are more electrically or magnetically active.

Work funded by the EC-FP6 (Grant No. STREP\_FP6-03321) and the Spanish DGI
(Grant Nos. FIS2006-12117-C04-01 and CSD2007-00041). We used the BSC-CNS and
CESGA Supercomputing Centers. We thank J.~Fontcuberta and F.~S\'anchez for
useful discussions.


\begin{thebibliography}{99}
\bibitem{fn:orbital} There are Co-based insulating spinels with considerable
  orbital moments, but the chances that such effects dominate a {\sl very
    large} ME response seem slim.
\bibitem{fn:extradets} We constrained the cell in-plane area by imposing
  $a(a+\delta_{1}-\delta_{2})=[a_{0}(1+\epsilon)]^{2}$, where $a_{0}$ is the
  computed pseudo-cubic lattice constant of the $R3c$ phase of bulk BFO. Thus,
  an in-plane skewing of the cell was allowed, but this distortion was found
  to be small. The epitaxial constraint was imposed in the response
  calculations by setting to infinity the corresponding elastic constants. The
  relaxed structures, as well as the $\boldsymbol{K}$, $\boldsymbol{C}$,
  $\boldsymbol{Z}$, $\boldsymbol{e}$ tensors and their derived quantities,
  were found to be essentially independent of the magnetic easy axis
  orientation.
\bibitem{fn:modesym} The obtained soft modes are fully symmetric, i.e., they
  preserve the $Cc$ symmetry of the {\sl R} and {\sl T} phases as compatible
  with isosymmetric {\sl R}$\leftrightarrow${\sl T} transitions. To analyze
  their character, we expressed them in the basis of the eigenmodes of
  $\boldsymbol{K}$ for cubic $Pm\bar{3}m$ BFO, where the FE ($\Gamma_{4}^{-}$)
  and AFD ($R_{4}^{+}$) components are distinct by symmetry. For these
  manipulations, and for the analysis of the character of the {\sl
    R}$\leftrightarrow${\sl T} transitions, we worked with relative atomic
  positions, thus ignoring cell changes.
\bibitem{fn:magsym} We checked the computed $\boldsymbol{\cal M}^{s}$'s lie
  along directions compatible with the magnetic
  class~\protect\cite{borovik06}. In the case of {\sl R}$^{II}$, the easy axis
  is [1$\bar{1}$0]$_{pc}$ (i.e., perpendicular to the symmetry plane of
  $Cc$) and the magnetic class $m'$, implying that $\boldsymbol{\cal M}^{s}$
  must be within the (1$\bar{1}$0)$_{pc}$ plane. We found $\boldsymbol{\cal
    M}^{s}$ rotates from [$\bar{1}$$\bar{1}$2]$_{pc}$ to
  [$\bar{1}$$\bar{1}$0]$_{pc}$ (see Fig.~2).
\bibitem{fn:minus6} For the {\sl R}~phase at $\epsilon$=$-$6\%, we
  were able to relax a structure that is a saddle point of the energy,
  being stable upon separate atomic and cell distortions (i.e.,
  $\boldsymbol{K}$ and the frozen-ion elastic tensor
  $\bar{\boldsymbol{C}}$ show no instability) but unstable upon a
  combined atom+cell perturbation (i.e., $\boldsymbol{C}$ shows an
  instability). The frozen-cell response of such a state is well
  defined.
\bibitem{fn:polarities} If $\boldsymbol{v}_{s}$ is the normalized
  soft-mode eigenvector, its dielectric (magnetic) polarity is
  $\boldsymbol{p}^{d}_{s} = \boldsymbol{v}_{s} \boldsymbol{Z}$
  ($\boldsymbol{p}^{m}_{s} = \boldsymbol{v}_{s}
  \boldsymbol{\zeta}$)~\protect\cite{wojdel09}.
\end{thebibliography}
\end{document}